\documentclass[11pt]{cernrep}
\usepackage{graphicx}
\usepackage{here}
\begin{document}
 \title{Gluon distributions in nuclei at small $x$: guidance from different
models}
\author{N. Armesto$^{a,b}$
and C. A. Salgado$^b$}
\institute{$^a$ Departamento de F\'{\i}sica, Universidad de C\'ordoba, Spain\\
$^b$ Theory Division, CERN, Switzerland}
\maketitle

The difference between the structure functions measured in nucleons and
nuclei \cite{Arneodo:1992wf} is a very important and well known
feature of the study of nuclear structure and nuclear collisions. At
small values of the Bjorken variable $x$ ($< 0.01$, shadowing
region), the structure function $F_2$ per nucleon turns out to be
smaller in nuclei than in a free nucleon. This shadowing corresponds to a
shadowing of parton densities in nuclei.
While at small $x$ valence quarks are of little importance and the
behaviour of the sea
is expected to follow that of $F_2^A$, the gluon distribution, which is
not an observable quantity, is badly
determined and
represents one of the largest uncertainties in computation of cross
sections both for moderate and large scales $Q^2$ in collinear
factorization \cite{Collins:gx}. For example, the
uncertainty in the determination of the glue for Pb at $Q^2\sim 5$ GeV$^2$ at
LHC for $y=0$, $x\sim m_T/\sqrt{s}\sim 10^{-4}\div 10^{-3}$, is a factor
$\sim 3$ (see Fig. \ref{compglue}),
which for the corresponding cross section in PbPb collisions
results in a factor $\sim 9$.

In this situation and while waiting for new experimental data to come
from lepton-ion \cite{Arneodo:1996qa,Abramowicz:2001qt,eacoll} or pA colliders,
the guidance from different theoretical models is of
uttermost importance to perform safe extrapolations from the region
where experimental data exist to those interesting for LHC studies.
Two different approaches to the problem have been essayed:
On the one hand, there exist models which try to explain the
origin of shadowing, usually in terms of multiple scattering (in the
frame where the nucleus is at rest) or parton interactions in the nucleus
wave function (in the
frame in which the nucleus is moving fast). On the other hand, other models
parameterize parton
densities inside the nucleus at some scale
$Q_0^2$ large enough for perturbative QCD to be applied reliably,
and then evolve these parton densities
using the DGLAP \cite{Dokshitzer:sg,Gribov:ri,Altarelli:1977zs}
evolution equations; in this way, the origin of the differences of
partons
densities in nucleons with respect to
nuclei is not addressed, but contained in the
parameterization at $Q_0^2$ which is obtained from a fit to experimental
data.

\section{Multiple scattering and saturation models}

The nature of shadowing is well understood
qualitatively:
In the rest frame of the nucleus,
the incoming photon splits, at small enough $x$,
into a $q\bar q$ pair
long
before reaching the nucleus, with a coherence length $l_c
\propto
1/(m_Nx)$ with $m_N$ the nucleon mass,
which at small enough $x$ becomes of the order of or greater than the nuclear
size.
Thus this $q\bar q$ pair interacts coherently with the nucleus
with typical hadronic cross sections, which results in absorption
\cite{Brodsky:1989qz,Barone:ej,Kopeliovich:1995yr,Armesto:1996id,Nikolaev:1990ja} (see
\cite{Armesto:2000zh} for a simple geometrical approach in this framework).
In this way nuclear
shadowing is a
consequence of
multiple scattering and is thus related with diffraction (see e.g. \cite{Capella:1997yv,Frankfurt:2002kd}).

Multiple scattering is usually formulated in the dipole model
\cite{Nikolaev:1990ja,Mueller:1994jq}, which is equivalent to $k_T$-factorization
\cite{Catani:1990eg} at
leading order. In this framework the $\gamma^*$-nucleus cross section is
expressed through the convolution of the probability of the transversal
or longitudinal $\gamma^*$ to split into a $q\bar q$ pair of transverse
dimension $r$ times the cross section for scattering of such dipole with
the nucleus. It is this dipole-nucleus cross section at fixed
impact parameter which
saturates (i.e. gets a maximum value allowed by unitarity), most
frequently by multiple scattering in the Glauber-Gribov approach
\cite{Armesto:2002ny,Armesto:2001vm,Huang:1997ii,Frankfurt:2002kd}. This dipole-target cross section is related
through a Bessel-Fourier transform to
the so-called unintegrated gluon distribution $\varphi_A(x,k_T)$ in 
$k_T$-factorization \cite{Andersson:2002cf,Armesto:2002ny,Armesto:2001vm}, which in turn can be related to the
usual collinear gluon density through
\begin{equation}
xG_A(x,Q^2)=\int_{\Lambda^2}^{Q^2} dk_T^2\  \varphi_A(x,k_T)
\label{gluglu}
\end{equation}
(with $\Lambda^2$ some infrared cut-off, if required)
although this identification is only true
for large $Q^2\gg Q_s^2$ \cite{Kovchegov:1998bi,Mueller:1999wm},
with $Q_s^2$ the saturation momentum
corresponding to a transverse length scale where the saturation of the
dipole-target cross section occurs.

Other formulations of multiple scattering do not use the dipole
formulation but relate shadowing with
diffraction \cite{Capella:1997yv,Frankfurt:2002kd} by Gribov theory. In this way diffraction in
lepton scattering on nucleons is related to nuclear shadowing. 

An equivalent explanation to multiple scattering in the
frame in which the nucleus is moving fast, is that gluon recombination
due
to the overlap of the gluon clouds from different nucleons, makes gluon
density
in nucleus with mass number
$A$ smaller than $A$ times that in a free nucleon \cite{Gribov:tu,Mueller:wy}: at small
$x$ the interaction develops over longitudinal distances $z\sim 1/(m_Nx)$
which become of the order of or larger than the nuclear size, leading to
the overlap of gluon clouds from different nucleons located in a
transverse area $\sim 1/Q^2$.
These studies have received great theoretical impulse with the
development of
semiclassical ideas in QCD and the appearance of non-linear equations
for
evolution in $x$ in this
framework (see
\cite{Mueller:2002kw,Iancu:2002xk} and references therein),
although saturation appears to be
different from shadowing \cite{Kovchegov:1998bi,Mueller:1999wm} (i.e. the reduction in the
number of gluons as defined in DIS is not apparent). In
the semiclassical framework the gluon field at saturation reaches a
maximum value and becomes proportional to the inverse QCD coupling
constant, gluon correlations are absent and a form for the dipole-target
cross section appears which leads to a geometrical scaling of $\gamma^*$-target
cross section; such scaling has been found in small $x$ nucleon data
\cite{Stasto:2000er}, and also in analytical and numerical solutions of the
non-linear equations \cite{Lublinsky:2001bc,Golec-Biernat:2001if,Armesto:2001fa,Iancu:2002tr}, but apparently the region where
this scaling should be seen in lepton-nucleus collisions has not been
reached yet in available experimental data \cite{Freund:2002ux}.
Apart from eA colliders \cite{Arneodo:1996qa,Abramowicz:2001qt,eacoll},
LHC will be the place to look
for non-linear effects.
However,
the consequences of such high density configurations may be masked in AB
collisions by other
effects, as those due to final state interactions. pA collisions would be, thus,
essential for this type of studies (see e.g. \cite{Dumitru:2002qt}). Moreover,
they would be required to fix the baseline for other studies in AB collisions
as QGP search and characterisation.
As a final comment, let us indicate that the non-linear terms in
\cite{Gribov:tu,Mueller:wy} are of a higher-twist nature and in this case the low density
limit recovers the DGLAP equations
\cite{Dokshitzer:sg,Gribov:ri,Altarelli:1977zs}, while the non-linear equations
for evolution in $x$
\cite{Mueller:2002kw,Iancu:2002xk} do not correspond to
any definite twist and their linear limit correspond to the BFKL equation
\cite{Fadin:cb,Balitsky:ic}.

Let us comment a little more on the difference between shadowing for gluons and
saturation. In saturation models
\cite{Mueller:2002kw,Iancu:2002xk}
saturation, defined by a maximum value of the gluon field or by the scattering
becoming black,
and shadowing for gluons, defined by the ratio $xG_A/(xG_N)$ being $<1$,
are apparently different phenomena, i.e. saturation does not necessarily
lead to shadowing for gluons \cite{Kovchegov:1998bi,Mueller:1999wm}.
Indeed, in the
framework of numerical studies of the
non-linear equations for small $x$ evolution, the unintegrated
gluon distribution turns out to be a universal function of just one variable
$\tau=k_T^2/Q_s^2$ \cite{Lublinsky:2001bc,Golec-Biernat:2001if,
Armesto:2001fa} (in \cite{Iancu:2002tr} this universality is analytically
shown to be
fulfilled up to $k_T^2$
much larger than $Q_s^2$), vanishing quickly
for $k^2_T>Q_s^2$; this scaling
appears also in the analysis of DIS experimental data on
nucleons \cite{Stasto:2000er} and
has been searched for in nuclear data \cite{Freund:2002ux}.
In nuclei, $Q_s^2$ increases
with increasing nuclear size, centrality and energy.
So, through the relation with the collinear gluon given by Eq. (\ref{gluglu}),
this scaling implies
that the integral gives the same value (up to logarithmic corrections if
a perturbative tail $\propto 1/k_T^2$ exists for large $k^2_T\gg Q_s^2$),
provided that the upper limit of the
integration domain $Q^2\gg Q_s^2$. Thus the ratio $xG_A/(xG_N)$ keeps equal to 1
(or approaching 1 as a ratio of logs if the perturbative tail exits)
with $x$ for $Q^2\gg Q_s^2$. So saturation or
non-linear evolution do not automatically guarantee the existence of
shadowing for gluons.
As
shadowing is in many models described in terms of
multiple scattering, this result
\cite{Kovchegov:1998bi,Mueller:1999wm} looks somewhat surprising.

\section{DGLAP evolution models}

On the other hand, a different approach is taken in
\cite{Eskola:1998iy,Eskola:1998df,Hirai:2001np,Indumathi:1996pb}: parton
densities inside the nucleus are parameterized at some scale
$Q_0^2\sim 1\div 5$ GeV$^2$ and then evolved using the DGLAP \cite{Dokshitzer:sg,Gribov:ri,Altarelli:1977zs}
evolution equations.
In this way, all nuclear effects on parton densities are included in the
parameterization at $Q_0^2$, which is obtained from a fit to experimental data.
The differences in the approaches come mainly from the sets of data used (e.g.
the use of Drell-Yan data or not) to constrain the parton distributions. In
these calculations the lack of experimental data makes the gluon to be badly
constrained at very small $x$; experimental data on the evolution of
$F_2^A$ with $\log{Q^2}$ give direct constrains \cite{Eskola:2002us} through
DGLAP evolution
but only for $x\geq 0.01$.
Please
refer to \cite{Eskola:2002us} and
references therein
for more discussions on this
kind of
models.

\section{Comparison between different approaches}
The results from different models usually depend on additional
semiphenomenological assumptions and often contradict each other. For
example,
concerning
the $Q^2$-dependence of the effect,
in \cite{Brodsky:1989qz,Barone:ej,Kopeliovich:1995yr,Armesto:1996id,
Nikolaev:1990ja} it is argued that $q\bar q$
configurations of a large dimension give the dominant contribution to
the
absorption, which results essentially independent of $Q^2$.
On the other
hand, in the gluon recombination approach of \cite{Gribov:tu,Mueller:wy} the absorption is
obtained as a clear higher-twist effect dying out at large $Q^2$.
Finally,
in the models \cite{Eskola:1998iy,Eskola:1998df,Hirai:2001np,Indumathi:1996pb}
which use DGLAP, all $Q^2$-dependence comes
from
QCD evolution and is thus of a logarithmic, leading-twist
nature.
Predictions (particularly for the gluon density) on the $x$-evolution
towards
small $x$ turn out to be very different.

\begin{figure}
\begin{center}
\includegraphics[width=12.5cm]{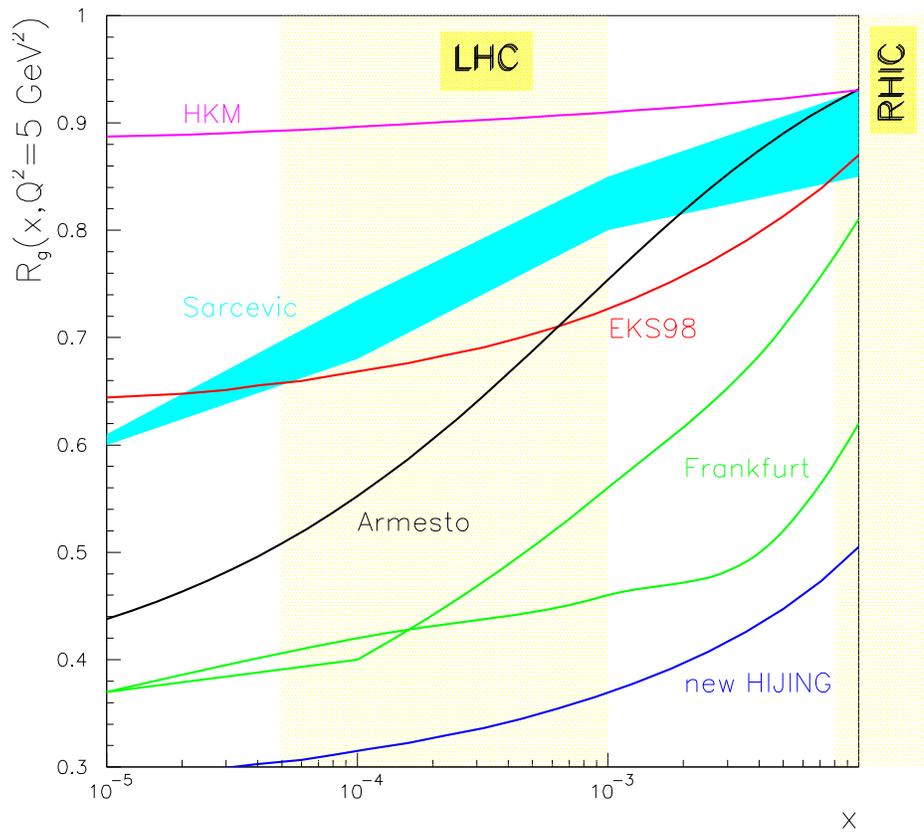}
\caption{Ratios of gluon distribution functions from different models at $Q^2$=5
GeV$^2$; HKM refers to the results from \protect{\cite{Hirai:2001np}},
Sarcevic to those from \protect{\cite{Huang:1997ii}}, EKS98 to those from
\protect{\cite{Eskola:1998iy,Eskola:1998df}}, Frankfurt to those from
\protect{\cite{Frankfurt:2002kd}}, Armesto to those from
\protect{\cite{Armesto:2002ny,Armesto:2001vm}} and new HIJING to those from
\protect{\cite{Li:2001xa}}. The bands represent the ranges of $x$
relevant for processes with a scale $Q^2=5$ GeV$^2$ at RHIC and LHC.}
\label{compglue}
\end{center}
\end{figure}

Let us compare between different approaches whose numerical results are
available (when considering an approach based on the dipole model as in
\cite{Armesto:2002ny}, one should keep in mind the
difficulties to identify at small and moderate $Q^2$
the integral of the unintegrated gluon
distribution with the ordinary gluon density, see Eq. (\ref{gluglu}) and
comments below it).
A comparison at $Q^2=5$ GeV$^2$
for the ratio of gluon densities in Pb over proton, can be found
in Fig. \ref{compglue}.
There it can be seen that the results of
\cite{Armesto:2002ny,Eskola:1998iy,Eskola:1998df,Hirai:2001np,Huang:1997ii} at
$x\simeq 10^{-2}$ (relevant for RHIC)
roughly coincide, while they are higher than
those of
\cite{Frankfurt:2002kd,Li:2001xa}; at $x\simeq 10^{-4}\div 10^{-5}$
(accessible at LHC)
the results of
\cite{Armesto:2002ny} become smaller
than those of \cite{Eskola:1998iy,Eskola:1998df,Hirai:2001np,Huang:1997ii}, get close to those of \cite{Frankfurt:2002kd} and
are still larger than those of \cite{Li:2001xa}. Apart from the
constraints coming from existing DIS experimental data on nuclei which
are very
loose for the glue at small $x$, in \cite{Eskola:1998iy,Eskola:1998df,Hirai:2001np}
the saturation of gluon shadowing comes mainly
from the initial condition for DGLAP evolution where
this saturation has been imposed, while \cite{Armesto:2002ny,Frankfurt:2002kd,Huang:1997ii} are
multiple scattering models.
In \cite{Li:2001xa} the behaviour of the glue, in the form of a
$Q^2$-independent parameterization, has been fixed at $x\sim 10^{-2}$ in
order to reproduce charged particle multiplicities in AuAu collisions
at RHIC.
In this approach the strongest gluon shadowing is obtained. However, it seems
to be in disagreement with existing DIS data \cite{Eskola:2002us}.
Additional caution has to be taken to compare results from multiple
scattering models with those
coming from DGLAP analysis \cite{Eskola:1998iy,Eskola:1998df,Hirai:2001np}: the ratios for the glue at some
moderate, fixed $Q^2$ and very small $x$ may
result smaller (e.g. in \cite{Armesto:2002ny})
than the ratios for $F_2$ at the same $x$, $Q^2$, which
might lead to
problems with leading-twist DGLAP evolution, see \cite{Eskola:2002us}.

\vskip 1cm
\noindent

\section*{ACKNOWLEDGEMENTS}
N. A. thanks CERN Theory Division, Departamento de F\'{\i}sica de
Part\'{\i}culas at Universidade de Santiago de Compostela, Department of
Physics at University of Jyv\"askyl\"a, Helsinki Institute of
Physics and Physics Department at BNL, for
kind hospitality during stays in which parts of this work have been
developed; he also acknowledges financial
support by CICYT of Spain under contract
AEN99-0589-C02 and by Universidad de C\'ordoba.
C. A. S. is supported by a Marie Curie fellowship of the European Community
programme TMR (Training and Mobility of Researchers) under the contract number
HPMF-CT-2000-01025.

\end{document}